\documentstyle[aps]{revtex}

\input{epsf}

\begin{document}
\author{D. Blume$^{(1)}$ and Chris H. Greene$^{(2)}$}
\title{
Three particles in an external trap: Nature of the complete $J=0$ spectrum}
\date{\today }
\address{$^{(1)}$Department of Physics, Washington State University, 
Pullman, WA 99164-2814, USA;\\
$^{(2)}$Department of Physics and JILA, University of Colorado, 
Boulder, CO 80309-0440, USA}
\maketitle

\begin{abstract}
\end{abstract}
Three bosonic, spin-polarized atoms in a spherical oscillator potential
constitutes the simplest nontrivial Bose-Einstein condensate (BEC).  The 
present paper develops the tools needed to understand the nature of the
complete $J=0$ energy spectrum for this prototype system, assuming a 
sum of two-body potentials.  The resulting spectrum
is calculated as a function of the two-body scattering 
length $a_{sc}$, which 
documents the evolution of certain many-body levels that evolve from 
BEC-type to molecular-type as the scattering length is decreased.  
Implications for the behavior of the condensate excited-state spectrum 
and for condensate formation and decay are elucidated.  
The energy levels evolve smoothly, even through the regime where 
the number of two-body bound states $N_b$ increases by 1,
and $a_{sc}$ switches from $-\infty$ to $+\infty$.
We point out the possibility of suppressing three-body
recombination by tuning the two-body scattering length to values that
are larger than the size of the condensate ground state.
Comparisons with mean-field treatments are presented.

\draft


\section{Introduction}
\label{introduction} 
Mean-field methods are commonly used to characterize
Bose-Einstein condensates (BECs), i.e., atomic alkali vapors
in an external trapping potential~\cite{dalf98}.
These approaches,
which account for the interaction between
particles through a mean-field term in an effective 
single-particle Hamiltonian,
accurately describe much of the dilute condensate
energetics.
However, these 
shape-independent $s$-wave scattering length approximations 
break down for strongly interacting
gases with
large interaction 
parameters~\cite{dalf98,braa97,lieb00,corn00,donl01}.

Very recent experiments by Greiner {\em{et al.}}~\cite{grei02,stoo02}
have entered this regime: A quantum phase 
transition from 
a superfluid to a Mott insulator was observed for 
ultracold atoms held in a three-dimensional
optical lattice.
Most relevant to our studies here,
the atoms are held in tight {\em{isotropic}} lattice sites
with an occupancy of 1-3 atoms per site.
The experiment thus realizes a strongly correlated many-body 
(and in a sense, few-body) 
quantum system with unprecedented
control of parameters.
In this paper, 
we present the quantal $J=0$ energy spectrum of three particles
under external isotropic confinement, and compare
with results obtained using mean-field theory.

The lowest order mean-field approximation [Gross-Pitaevskii (GP)
or Hartree-Fock (HF) equation]~\cite{dalf98,pita61,gros61,gros63,esry97} 
and also higher order approaches such as 
second order perturbation theory~\cite{morg00,davi01}  
treat the metastable BEC
state of the trapped alkali vapor as if it was the 
``true'' ground state of the system.
This means that
these approaches cannot 
describe recombination processes or atom-loss processes,
since coupling to the true (solid-like/liquid-like)
ground state is neglected completely.
Attempts to include additional ``empirical'' terms
in these mean-field equations,
which account for ``reaction processes'', exist, and have 
been able to describe some experimental findings 
successfully~\cite{kaga98,duin01,sait01}. However,
the ad hoc nature of some of these empirical terms 
is less than satisfactory.

The nonperturbative quantal calculations permit us to circumvent these 
problems. These calculations come at a price, namely, our initial restriction
to only $N=3$ particles with total angular momentum $J=0$
in the trap.
For this ``simple'' system, however, we are able to include 
all channels in the calculation,
molecular three-body bound state channels, 
diatom plus atom channels, and metastable
gaseous-like channels~\cite{esry99b}. 

This paper focuses on regimes where 
standard mean-field treatments break down.
For large positive $a_{sc} \rightarrow +\infty$, the mean-field treatment results in
unphysically large diverging energies.
In contrast, the three-body energies change smoothly
in the pole region, where
the scattering length $a_{sc}$ changes from $-\infty$
to $+\infty$ and
the number of two-body $s$-wave bound states $N_b$ 
increases by $1$.
Additionally, our
three-body study implies
that the excitation frequencies 
are well behaved around $|a_{sc}| \rightarrow \infty$. 
 
The mean-field treatment also exhibits limitations
in the parameter range around the negative 
critical scattering length $a_{cr}$,
where $a_{cr}$ is
defined through the instability criterion
$(N-1)a_{cr}/a_{ho}=-0.575$ derived from the 
mean-field equation~\cite{esry97,dodd96}.
The lowest gaseous energy level (i.e., the BEC ground state)
changes its character from metastable to molecular 
around $a_{cr}$, while some of the higher 
lying BEC-like states remain almost completely unaffected by 
criticality. 
In contrast, the mean-field equation simply has no solution
 for $a_{sc}$ lower than the critical value.

Three-particle studies have been used previously to shed 
light on the many-body physics of BECs.
In 1996, Esry {\em et al.}~\cite{esry96a} explored the role of the 
two-body scattering length 
in BECs.
More recently, three research groups explained 
experiments regarding the three-body recombination rate in many-atom
BECs via three-body calculations that include channel 
coupling~\cite{esry99b,niel99,beda00,kart01,braa01}.
Three-body studies have also been used
to test the validity of shape-independent two-body
interaction potentials~\cite{esry99}.

The studies presented here are based on time-independent Schr\"odinger
wave mechanics.
The energy levels show several avoided
crossings when plotted as functions of $a_{sc}$ 
(see Section~\ref{energies}); nevertheless, dynamical 
behavior such as Landau-Zener-type
branching ratios 
or recombination rates remains beyond the scope of this paper.
Studies along these lines will be a natural continuation
of the static studies presented here. 

Section~\ref{system} describes our three-body system, and 
outlines the formalism for its investigation. Section~\ref{energies} presents
quantum mechanical three-body energetics, and Section~\ref{meanfield}
compares with results derived from mean-field treatments. 
Finally, Section~\ref{summary} summarizes our main results.

\section{The three-particle system}
\label{system} 
The Schr\"odinger equation for three mass-$m$ particles in a spherical external
potential with trapping frequency $\nu_{ho}=\omega_{ho}/(2 \pi)$
reads 
\begin{eqnarray}
	\left[ \sum_{i=1}^3 -\frac{\hbar^2}{2m} \nabla^2_{\vec{r}_i} +
	\sum_{i=1}^3 \frac{1}{2} m \omega_{ho}^2 {r}_i^2 +
	\sum_{i<j}^3 V(r_{ij}) \right]
	\Psi(\vec{r}_1,\vec{r}_2,\vec{r}_3) = 
	E \Psi(\vec{r}_1,\vec{r}_2,\vec{r}_3),
	\label{eq_cartesian}
\end{eqnarray}
where $V(r_{ij})$ denotes a two-body interaction potential and
$r_{ij} = | \vec{r}_j-\vec{r}_i |$.
$\vec{r}_i$ is the position vector of atom $i$ relative
to the center of the trap.
Restriction to states of zero total angular momentum ($J=0$) reduces this
nine-dimensional problem to six dimensions.
The dimensionality of the problem can be reduced further
by a transformation to Jacobi coordinates,
$\vec{X}_{cm} = (\vec{r}_1 + \vec{r}_2 + \vec{r}_3)/3$,
$\vec{\rho}_{12}= \vec{r}_2 - \vec{r}_1$, and
$\vec{\rho}_{12,3} = \vec{r}_3 - (\vec{r}_2+\vec{r}_1)/2$~\cite{esry99}.
This transformation decouples the center-of-mass motion,
\begin{eqnarray}
	\left( -\frac{\hbar^2}{2M} \nabla^2_{\vec{X}} +
	\frac{1}{2} M \omega_{ho}^2 X^2 \right)
	\phi_{klm}(\vec{X}) =
	E_k^{CM} 	\phi_{klm}(\vec{X}),
	\label{eq_cm}
\end{eqnarray}
from the relative (internal) motion,
\begin{eqnarray}
	\left[
	-\frac{\hbar^2}{2\mu }\frac{\partial ^{2}}{\partial R^{2}}+
	\frac{1}{2} \mu \omega_{ho}^2 R^2 +
	\frac{\hbar^2}{2\mu R^{2}} 
	\left( \Lambda^2+ \frac{15}{4} \right)+
	\sum_{i<j}^3V(r_{ij}) \right]
	\psi_n(R,\vartheta,\varphi) = 
	E^{int}_n\psi_n(R,\vartheta,\varphi).
	\label{eq_internal}
\end{eqnarray}
Here, $M=3m$ is the total mass, and $E_k= (k+3/2) \hbar \omega_{ho}$
with $k=0,1,2,\cdots$ the energy related to the center-of-mass motion.
The corresponding eigenfunctions $\phi_{klm}$ 
are simply the three-dimensional isotropic oscillator
solutions.
Eq.~(\ref{eq_internal}) represents the 
Hamiltonian describing the relative particle motions
in terms of three hyperspherical coordinates ($R$, $\vartheta$, 
$\varphi$) rather than the Jacobi coordinates $\vec{\rho}_{12}$ and 
$\vec{\rho}_{12,3}$~\cite{esry99,mace68,lin95,zhou93}.
Our definition of the hyperradius $R$, and the two hyperangles 
$\vartheta$ and $\varphi$ is based on the
democratic coordinate system of Whitten-Smith~\cite{whit68},
i.e., $\mu R^2= m \rho_{12}^2/2+ 2m \rho_{12,3}^2/3$.
Owing to the identical boson character of the atoms, the angular range
can be restricted to $\varphi \in [ 0, \pi / 6]$ and
$\vartheta \in [ 0, \pi / 4]$.
$\mu = m/\sqrt{3}$ denotes the reduced mass of the
three-body system, while $\Lambda^2$ denotes
the ``squared grand angular momentum'' operator~\cite{aver89}.
Here we rescale the wave function $\psi(R,\vartheta,\varphi$) 
by $R^{5/2}$, i.e. 
$\Psi(\vec{r}_1,\vec{r}_2,\vec{r}_3) = 
R^{-5/2} \psi(R, \vartheta, \varphi) \phi_{klm}(\vec{X})$,
in order to eliminate first derivatives from the 
hyperradial kinetic
energy operator.

Eq.~(\ref{eq_internal}) is solved here using
two different, but related approaches: 
a coupled-adiabatic-channels approach and
a strict adiabatic approximation.
The coupled-adiabatic-channel calculation can in principle be made
exact.  The strict adiabatic approximation can be viewed as an 
``incomplete'' coupled-adiabatic-channel
calculation, truncated to one channel only, which 
neglects all off-diagonal coupling elements.
It is important to keep in mind that the strict adiabatic 
approximation cannot describe recombination processes.

To solve Eq.~(\ref{eq_internal}) we expand $\psi_n(R,\vartheta,\varphi)$
into radial wave functions $F_{\nu n}(R)$ and 
a set of complete, orthonormal angular channel functions
$\Phi_{\nu}(R;\vartheta,\varphi)$.
The basis functions $\Phi_{\nu}$, which depend parametrically
on $R$, are solutions
of the partial differential equation
\begin{eqnarray}
	\left[
	\frac{\hbar^2}{2\mu R^{2}} \left(\Lambda^2+ \frac{15}{4}\right)+
	\sum_{i<j}^3V(r_{ij}) \right]
	\Phi_{\nu}(R;\vartheta,\varphi) = 
	U_{\nu}(R) \Phi_{\nu}(R;\vartheta,\varphi).	
	\label{eq_angular}
\end{eqnarray}
We refer to the eigenvalues $U_{\nu}(R)$ as adiabatic potential
curves.
Inserting the expansion $\psi_n = \sum F_{\nu n} \Phi_{\nu}$ 
into the Schr\"odinger equation for the internal motion
[Eq.~(\ref{eq_internal})]
results in an infinite set of coupled ordinary differential equations,
\begin{eqnarray}
	\left( -\frac{\hbar^2}{2\mu }\frac{d^{2}}{dR^{2}}+
	\frac{1}{2} \mu \omega_{ho}^2 R^2 +
	U_{\nu }(R)+Q_{\nu \nu}(R)-E_{n}^{int}\right) F_{\nu n}(R) 
	&=&  \label{eq_radial} \\
	&&-\sum_{\nu' \neq \nu }Q_{\nu \nu' }(R)F_{\nu' n}(R)+
	\sum_{\nu' \neq \nu
	}P_{\nu \nu' }(R)\frac{dF_{\nu' n}(R)}{dR}, \nonumber
\end{eqnarray}
where $P_{\nu \nu' }$ [$Q_{\nu \nu' }$] 
are angular coupling matrix elements
involving the first [second] derivative of $\Phi_{\nu}$
with respect to $R$~\cite{blum00a}.
We refer to the solution of 
Eq.~(\ref{eq_radial}) as the coupled-adiabatic-channel solution.
The strict adiabatic approximation follows immediately 
by neglecting coupling elements between different channels 
[$\equiv$ setting the right hand side of Eq.~(\ref{eq_radial}) to zero].
Another variant of the strict adiabatic approximation 
additionally neglects 
the diagonal element
$Q_{\nu\nu}$, and is typically referred to
as the hyperspherical
Born-Oppenheimer approximation.
In the following, we use these two variants of
the strict adiabatic approximation interchangably,
and denote the eigenvalues by $E_{\nu n}^{int}$.

The outlined formalism splits the solution of 
the Schr\"odinger equation for the internal motion [Eq.~(\ref{eq_internal})] 
into two steps: 
{\it i)} solution of Eq.~(\ref{eq_angular}) 
(here via a two-dimensional B-spline code),
and
{\it ii)} solution of the one-dimensional 
coupled equations, Eq.~(\ref{eq_radial}).
The former is numerically more challenging
due to its higher dimensionality, and coincides with
determining the bound and continuum channel functions
$\Phi_{\nu}$
of three particles interacting via a sum of two-body potentials $V(r)$
[the trapping potential does not enter Eq.~(\ref{eq_angular})].
 
For a vanishing interaction potential,
$V(r)=0$,
the eigenfunctions $\Phi_{\nu}$
of Eq.~(\ref{eq_angular}) reduce to 
Gegenbauer polynomials~\cite{aver89}
with
eigenvalues 
$U_{\nu}(R) = \hbar^2 \frac{\lambda(\lambda+1)+15/4}{2 \mu R^2}$,
$\lambda = 0,4,6,\cdots$,  
where the $\lambda=2$ state is forbidden due to symmetry constraints, and the
eigenvalues with $\lambda=12$ and $\lambda \ge 16$ are doubly-degenerate.
Now consider a non-vanishing 
two-body interaction potential $V(r)$.
Compared to the $V=0$ potential, 
the non-vanishing $V$ introduces 
three-body bound states at small hyperradii $R$, accounting for
the short range physics of $V(r)$, and also modifies the 
potential curves at large $R$, reflecting the non-zero
$s$-wave scattering length $a_{sc}$.
Additionally, a non-vanishing $V$ lifts the degeneracy of 
the $U_{\nu}(R)$ potential curves.

To illustrate this behavior,
Fig.~\ref{fig_pot1} shows the sum of the trapping potential,
$V_{trap}=\mu \omega_{ho}^2 R^2/2$ with $\nu_{ho} = 78$kHz,
and the
adiabatic potential curves,
$U_{\nu}(R)$, 
for a two-body model interaction potential (see Sec.~\ref{energies}) with 
$s$-wave scattering length $a_{sc}=228$a.u. and 
two two-body $s$-wave bound states, $N_b=2$ [$m = m(^{87}$Rb)].
Length and energy are expressed in
oscillator units (length unit $a_{ho}=\sqrt{\hbar/(m \omega_{ho})}$,
energy unit $\hbar \omega_{ho}$).
The two lowest potential curves with
$\nu = 0$ and $1$ (upper panel, dashed lines) describe
three-body bound state physics,
and approach the two-body $s$-wave binding energies in the
absence of a trapping potential.
The higher lying potential curves (solid lines)
describe metastable BEC physics, and would correspond
to continuum states, in the absence of a 
trapping potential. The $\nu = 2$
curve (thick solid line) shows a double minimum structure
with minima at $R \approx 0.2 a_{ho}$ and $R \approx 2.5 a_{ho}$,
which are separated by a potential barrier.
The lowest metastable state ``lives'' in the minimum at larger 
$R$.
The lower panel of Fig.~\ref{fig_pot1} shows an
enlargement of the gaseous region together with
the effective potential curves 
for a vanishing interaction potential (dotted lines).
Significant deviations between the solid and the dotted curves 
(non-vanishing and vanishing two-body potential, respectively) are visible.

As clearly demonstrated in Fig.~\ref{fig_pot1},
to
properly describe the metastable states,
one must
solve Eq.~(\ref{eq_angular}) for hyperradii $R$
as large as $10 a_{ho}$, corresponding roughly to several $10^4$a.u.
(depending somewhat on $\nu_{ho}$). 
In our calculation, we
determine the adiabatic potential curves $U_{\nu}$ and the 
coupling matrix elements $P_{\nu \nu'}$ and 
$Q_{\nu \nu'}$ out to $R \approx 10-100 a_{sc}$ using
a two-dimensional B-spline code, 
and then extrapolate to larger $R$.
For the continuum potential curves we follow
a second independent approach suggested by Nielsen and Macek~\cite{niel99}.
For large $R$, they 
recognise the dependence of $U_{\nu}$ on $a_{sc}$ only,
and derive a nearly analytic expression for $U_{\nu}$.
Note, Nielsen and Macek's formula does not account for those
potential curves,
which simply coincide with the ``unperturbed'' eigenvalues
$\hbar^2 \frac{\lambda(\lambda+1)+15/4}{2 \mu R^2}$, 
$\lambda=12,16,18,\cdots$ (see above).
The nearly analytical expressions~\cite{niel99} 
are in excellent agreement with 
our continuum curves $U_{\nu}$ calculated through numerical solution 
(and extrapolation) of 
Eq.~(\ref{eq_angular}).

In passing, we note an additional nice feature of the outlined formalism.
Suppose that we wish to study the energetics
of three particles in an external 
spherical potential as a function of the trapping frequency
$\nu_{ho}$.
For a given interaction potential $V(r)$, it is only necessary to solve 
the numerically most demanding Eq.~(\ref{eq_angular}) once.
$\omega_{ho}$ then enters the simple one-dimensional
Eq.~(\ref{eq_radial}) simply through an additive term, which 
trivially determines the energy spectrum 
versus $\omega_{ho}$ .

\section{The quantal energy spectrum}
\label{energies} 
This section presents the internal energies 
$E^{int}$ for three bosonic particles with $J=0$ 
in an external trap 
for a two-body model 
potential of the form $V(r)=d \cosh^{-2}(r/r_0)$.
This is convenient because 
the two-body $s$-wave eigenenergies and eigenfunctions
can be determined analytically~\cite{land77}.
Furthermore, we determined an analytical expression for the 
energy-dependent $s$-wave scattering length,
\begin{eqnarray}
a_{sc}(k)= \frac{1}{ik}
      \frac{1+\rho}{1-\rho},
  \label{SechScattLength1}
\end{eqnarray}
where
\begin{eqnarray}
\rho = -\frac{_{2}F_{1}(ik-s,1+s+ik;1+ik;\frac{1}{2})}{%
2^{ik}\;_{2}F_{1}(-s,1+s;1-ik;\frac{1}{2})},
  \label{SechScattLength2}
\end{eqnarray}
where $k=\sqrt{m E/ \hbar^2}$, and
$s = -{1\over2}+{1\over2}\sqrt{1+4 m d r_0 ^2 / \hbar^2}$.
$_2F_1$ denotes the usual hypergeometric function.
The zero-energy $s$-wave scattering length $a_{sc}$ is
then simply given by $a_{sc}=\lim_{k \rightarrow 0} a_{sc}(k)$.
While $r_0$ is fixed at $r_0=55$a.u.
throughout our calculations, 
$d$ is varied to change the number of
two-body $s$-wave bound states $N_b$ and the 
two-body $s$-wave scattering length $a_{sc}$
[$m=m(^{87}$Rb) throughout the rest of the paper].
We perform three-body calculations for 70 different $d$ values,
which translates into a range of two-body scattering lengths $a_{sc}$
from $-10^4$a.u. to $10^4$a.u..
To keep the coupled-adiabatic-channel calculations tractable we
restrict $d$ in this study to values such that $N_b = 1$ or $2$.

Consider a
vanishing interaction potential, $d=0$ (ideal gas) first.
For this system, the strict adiabatic approximation is exact,
and 
$E^{int}_n=3,5,7,\cdots \hbar \omega_{ho}$.
Introduction of a non-vanishing interaction potential
shifts these gaseous energy levels according
roughly to the value of $a_{sc}$, lifts the degeneracy, and additionally
gives rise to three-body bound states.
Figure~\ref{fig_energy1} shows the internal energies 
$E_{\nu n}^{int}$ (solid lines) of the gaseous-like 
states for $\nu_{ho}=78$kHz calculated via the
strict adiabatic approximation
for various scattering lengths $a_{sc}$.
To accomodate a large range of interaction parameters
$(N-1) a_{sc}/a_{ho}$~\cite{esry97}, Fig.~\ref{fig_energy1} shows the
internal energies as a function of 
$[\mbox{arctan}(a_{sc}/a_{ho})]/\pi$.
To additionally be able to plot results for systems with $N_b=1$ and
$N_b=2$ two-body bound states, respectively,
the integer part of the abcissa is chosen to be equal
to $-N_b$. This scaling procedure results in
abcissa values of
$[-2.5,-0.5]$.
We refer to the interval $[-0.5,-1.5]$ as the
``first cycle'' ($N_b=1$), and to the interval $[-1.5,-2.5]$ 
as the
``second cycle'' ($N_b=2$). 
In this representation, $a_{sc}=0$ corresponds to 
$[\mbox{arctan}(a_{sc}/a_{ho})]/\pi=-2.0$ and $-1.0$.

To interpret this complicated energy level 
structure, concentrate on
six different scattering length values labeled by A--F in
Fig.~\ref{fig_energy1}. Figure~\ref{fig_pot2} shows the sum of
the corresponding adiabatic potential curves 
and the trapping potential, 
$U_{\nu}+V_{trap}$, where 
$V_{trap}(R)=0.5 \mu \omega_{ho}^2 R^2$ (solid lines):

Panel A corresponds to a large positive
scattering length, $a_{sc}=2138$a.u..
The lowest BEC-like curve ($\nu=2$) is highlighted as a thick
solid curve.
The lowest energy level lying within
this potential curve, $E_{20}^{int}$, has an energy close to
$5 \hbar \omega_{ho}$, and the excited levels 
($E_{2n}^{int}$, $n>0$) lie 
about
$2 \hbar \omega_{ho}$ higher each (see Fig.~\ref{fig_energy1}).
We refer to the energy levels ``located'' within the same potential 
curve (identical $\nu$, but having a different number of hyperradial 
nodes $n$) as a family.
The lowest energy of the next higher BEC-like curve is 
$E_{30}^{int} \approx 7 \hbar \omega$,
and again, the excited levels ($E_{3n}^{int}$, $n>0$) lie 
about
$2 \hbar \omega_{ho}$ higher each. 
To guide the eye, Fig.~\ref{fig_energy1} displays dotted curves that 
represent the effective potential curves in the absence of any 
two-body interaction 
potential.

Panel B corresponds to a much smaller
scattering length than in panel A, $a_{sc}=103$a.u.. 
This panel shows a series of crossings at energies around 
$20 \hbar \omega_{ho}$,
indicating
a $d$-wave shape resonance.
As the scattering length decreases slightly, these crossings 
move to lower energies,
until they finally vanish at $a_{sc}=97$a.u..
The scattering length $a_{sc}=97$a.u.
corresponds to a well depth $d$ of the two-body
potential $V$ at which $V$ first supports a $d$-wave two-body bound state.
To calculate energy levels in the presence of a
$d$-wave shape resonance we follow the potential curves that cross diabatically.
Note that the potential curves approach the $V(r_{ij})=0$
curves (i.e., the dotted lines) as $a_{sc}$ approaches zero. 
Accordingly, the energy levels approach those of the ideal inhomogeneous gas.

Panel C shows the effective potential curves (solid lines)
for vanishing scattering length, $a_{sc}=0$ (however, non-vanishing
$V$),
in which case the potential curves are almost identical to 
the ones for vanishing $V$ (dotted lines).
Small differences occur in the inner wall region reflecting the existence 
of molecular states at small $R$. These molecular states
at small $R$ are, however, not shown in Fig.~\ref{fig_pot2}.

Panels D, E, and F correspond to negative
scattering lengths. 
The inflection of the thick solid line at $R \approx 0.5 a_{ho}$ 
in panel $D$ indicates the 
first signature of a decaying condensate. The interaction
parameter for this system is $(N-1)a_{sc}/a_{ho} = -0.085$.
In panel E, the barrier of the lowest gaseous-like curve,
which separates the BEC minimum from the energetically
much deeper molecular minimum, 
has almost vanished;
here, $(N-1)a_{sc}/a_{ho} = -0.356$. 
Panel F shows a large negative scattering length case,
$a_{sc}=-1970$a.u..
The thick solid curve corresponds to a situation
in which the gaseous-like state (condensate) is no longer stable.
Note the similarity of the potential curves for large
positive and large negative $a_{sc}$ (panel A and F).
This similarity reflects the smooth dependence of  
$E^{int}_{\nu n}$ as $|a_{sc}|$ goes through infinity (see below).

In addition to the aforementioned issues, four more
characteristics are important to the energy level 
scheme shown in Fig.~\ref{fig_energy1}:
{\em (i)} At $a_{sc}=0$, corresponding to
$[\mbox{arctan}(a_{sc}/a_{ho})]/\pi=-1$ and $-2$, 
the positive energy levels $E^{int}_{\nu n}$
lie very close to those of the ideal gas.
{\em (ii)} The energy level with energy $E^{int} = 15 \hbar \omega_{ho}$
is unchanged over the two cycles shown. This energy level 
originates in the ``unperturbed''
potential curve $\hbar^2 \frac{\lambda(\lambda+1)+15/4}{2 \mu R^2}$,
$\lambda=12$ (see Section~\ref{system}),
and therefore coincides with an ideal gas level.
{\em (iii)} The internal energy levels in the first cycle ($N_b=1$) and
in the second cycle ($N_b=2$) behave very similarly, indicating 
that the energies are, to a good approximation, independent
of the number of two-body $s$-wave bound states $N_b$. 
{\em (iv)} Notice in particular
that the energy levels $E^{int}_{\nu n}$ change 
smoothly as $|a_{sc}|$ passes through infinity.

Note as well that
 families of plunging energy
levels with molecular character are present, 
which to simplify our presentation 
are shown only partially in Fig.~\ref{fig_energy1}.
Consider the lowest gaseous-like energy level of the first cycle,
i.e. with $N_b=1$.
As the gaseous-like state becomes unstable while $a_{sc}$ decreases  
 this energy level decreases through zero
(at $[\mbox{arctan}(a_{sc}/a_{ho})]/\pi \approx -1.2$), 
which reflects the transition of this BEC state to one of 
molecular character.
The other energy levels of this family also change character from gaseous-like
to molecular-like as the two-body potential well depth
$d$ becomes deeper. 
For clarity, Fig.~\ref{fig_energy1}
shows these energy levels only for 
$[\mbox{arctan}(a_{sc}/a_{ho})]/\pi > -1.75$.
For each additional two-body bound state, a
``manifold'' of plunging energy levels is present, reflecting the fact that the
number of three-body states with negative energy 
increases far more rapidly than those for the two-body system.

So far we have discussed results for a trap with trapping 
frequency
$\nu_{ho}=78$kHz.
In addition, we performed calculations for trapping frequencies 
that are 10 times
larger, and also 10 or 100 times smaller.
Since $a_{ho}=\sqrt{\hbar/(m \omega_{ho})}$,
the interaction parameter $(N-1)a_{sc}/a_{ho}$ scales as $\sqrt{\omega_{ho}}$.
Adopting the same two-body interaction potentials as before,
a trapping frequency smaller than $\nu_{ho}=78$kHz
therefore leads to a narrower 
$[\mbox{arctan}(a_{sc}/a_{ho})]/\pi$ interval than
shown in Fig.~\ref{fig_energy1}.
As an example, Fig.~\ref{fig_energy2} 
shows the internal energies $E_{\nu n}^{int}$ for three particles in
a trap with $\nu_{ho}=780$Hz and $N_b=2$ (solid lines). 
A few of the internal energies 
at large negative scattering length,
corresponding to $[\mbox{arctan}(a_{sc}/a_{ho})]/\pi \approx -2.08$, are
indicated by diamonds. These data points 
are not connected by solid lines since in this region our
grid in $a_{sc}$ is too coarse to allow for interpolation.

To summarize, our microscopic studies 
of the energetics of three confined particles 
as a function of $\nu_{ho}$ reveal
that the overall behavior of the energy level scheme
(as shown in Fig.~\ref{fig_energy1}) remains unchanged as $\nu_{ho}$
either decreases or increases. 
As discussed 
in the previous paragraph, however,
$\nu_{ho}$ determines the range of the interaction parameter. 
Specifically, the behavior of the energy levels 
around $a_{sc} \approx a_{cr}$ 
depends on $\nu_{ho}$.
Figure~\ref{fig_energy1} indicates a 
gradual decrease of the condensate energy as the scattering length
decreases through the region
$[\mbox{arctan}(a_{sc}/a_{ho})]/\pi \approx -2.1$.
As $\nu_{ho}$ is decreased, the separation between the molecular-like
region at small $R$ and the gaseous-like region at large $R$ becomes more
pronounced, and the energy decrease
(change from gaseous-like to molecular-like) takes place over a 
smaller range of scattering lengths.
In other words, the potential barrier (shown in panel D and E of
Fig.~\ref{fig_pot2} for $\nu_{ho}=78$kHz) spatially 
separates the molecular-like
and gaseous-like region more for small
$\nu_{ho}$ than for large $\nu_{ho}$.

The internal energies discussed in this section have been calculated within
the strict adiabatic approximation.
To assess the accuracy of this approach we additionally
performed coupled-adiabatic-channel calculations.
Figure~\ref{energy_cc} shows the energies resulting 
from the coupled-adiabatic-channel
calculation including four channels (pluses) together with those 
resulting from the
adiabatic calculation (dotted, solid, and dashed lines)
for $\nu_{ho}=78$kHz. 
Figure~\ref{energy_cc} indicates excellent agreement 
between these two sets of calculations
for small $a_{sc}$
(left hand side of the figure), however, some differences occur at larger 
$a_{sc}$ (right hand side of the figure).
To be more specific, the plunging molecular levels (dotted lines)
and the BEC levels in the lowest gaseous adiabatic potential curve
(solid lines) seem to repell each other at large $a_{sc}$ 
due to channel coupling.
In summary,
for our time-independent studies here, the 
strict adiabatic approximation describes the main characteristics 
of the energy level pattern as shown in 
Figs.~\ref{fig_energy1} and \ref{fig_energy2}
with sufficient accuracy. 
To go beyond this static picture, however, the coupling between 
channels must be considered.

Studies including channel coupling~\cite{esry99b,esry96a} 
show that the primary recombination
mechanism
is related to the behavior of the adiabatic potential curves 
and the corresponding
coupling elements at hyperradii $R \approx 2-3a_{sc}$.
Consider panel A of Fig.~\ref{fig_pot2} 
with $a_{sc}=2138$a.u. and $\nu_{ho}=78$kHz.
For this situation, the criterion $R \approx 2-3a_{sc}$
translates into 
$5.8-8.8a_{ho}$.
At these hyperradii, the lowest lying gaseous $V_{trap}+U_{\nu}$ curve
reaches $\approx 10-20 \hbar \omega_{ho}$, compared to the lowest
gaseous energy level of $\approx 5.3 \hbar \omega_{ho}$.
Thus, we speculate that we have entered a regime
where recombination processes are largely 
suppressed due to the strong confinement. 
It would be interesting to probe this regime experimentally.

\section{Comparison with Mean-Field Treatments}
\label{meanfield}
This section connects our microscopic studies
with results obtained by mean-field treatments.
Consider the HF treatment~\cite{esry97}, which 
results in an equation identical to the GP
equation, except for a change from $N$ to $N-1$ in the interaction parameter.
The HF treatment does not take advantage of the  
decoupling of the center-of-mass
and the internal motion, and determines an approximate value 
for the total energy $E_{HF}$ as well as the 
orbital energy or chemical potential.
To compare our internal three-body energies $E_{\nu n}^{int}$ (solid lines
in Figs.~\ref{fig_energy1} and \ref{fig_energy2})
with the HF energy $E_{HF}$, we subtract
the exact energy associated with the lowest center-of-mass motion,
$E^{CM}_0=1.5\hbar \omega_{ho}$, from 
$E_{HF}$, and refer to this quantity as internal HF energy $E_{HF}^{int}$.
The approximate HF treatment 
could either lead to an error
in $E^{int}_{HF}$ or in $E^{CM}_{HF}$,
or in both these quantities at the same time, and
the subtraction of $E^{CM}_0$ from $E_{HF}$ is therefore somewhat
artificial~\cite{blum01a}.
Figures~\ref{fig_energy1} and \ref{fig_energy2} show the 
resulting internal
HF energy, $E_{HF}^{int}$, as dotted lines.

For $\nu_{ho}=780$Hz, the internal HF energy $E_{HF}^{int}$
(dotted lines in Fig.~\ref{fig_energy2}) agrees favorably with the 
lowest gaseous-like internal three-body energy (solid line).
As discussed in Sec.~\ref{energies}, 
Fig.~\ref{fig_energy2}
includes only a small range of interaction parameters $(N-1)a_{sc}/a_{ho}$. 
Thus, the good agreement between $E_{\nu n}^{int}$ and $E_{HF}^{int}$
is not surprising.

For comparison, Fig.~\ref{fig_energy1} shows the energy levels for
a 100 times larger trapping frequency, namely $\nu_{ho}=78$kHz.
This figure indicates two regions in which 
$E_{HF}^{int}$ (dotted line) deviates from the 
lowest many-body energy level $E^{int}_{\nu n}$ having gaseous character, 
the region around
$a_{cr}$ and 
the large $a_{sc}$ region.
For $a_{sc} \le a_{cr}$, the HF equation does not have a solution.
This mathematical fact is typically interpreted as ``decay
of the condensate''~\cite{dalf98,dodd96}. The three-body treatment 
reveals that the decay of the condensate is linked to the
disappearance of a potential barrier (see, e.g., panels D and E
of Fig.~\ref{fig_pot2}),
which separates the molecular region at small $R$ from the gaseous-like 
region at large $R$.
When the potential barrier has disappeared,
the states associated with 
this potential curve have purely molecular character.
Thus, the  
energy levels in this potential curve change dramatically in the vicinity
of $a_{cr}$, whereas most of the energy levels
belonging to other families change slowly around $a_{sc}$.

Our interpretation of the behavior of a BEC in the vicinity
of $a_{cr}$ agrees with that suggested by Bohn~{\em{et al.}}
for an arbitrary number of particles~\cite{bohn98}.
Within the $K$-harmonic approximation,
Bohn {\em{et al.}}~\cite{bohn98} describe
the interaction between each
pair of particles through a shape-independent 
delta-function potential, and make some approximations
about the behavior at small hyperradii, where the $\delta$-function
potential leads to divergences. These approximations are 
confirmed to be well
justified, based on our calculations with 
shape-dependent two-body interaction potentials,
so that no additional approximations at small $R$
need to be made.
Taken together, these two studies highlight
the usefulness of the hyperspherical radius coordinate in 
interpreting dynamics of a BEC (see also~\cite{hart00}).

Figure~\ref{fig_energy1} also indicates discrepencies between $E_{HF}^{int}$ 
and $E^{int}_{\nu n}$ at large $|a_{sc}|$.
For $a_{sc} \rightarrow \infty$,
$E_{HF}^{int}$ diverges and approaches unrealistically large
values, whereas the three-body energies $E_{\nu n}^{int}$
behave smoothly.
Recall that the sign change of the scattering length from $-\infty$
to $+\infty$ (pole region)
corresponds to the appearance of 
an additional two-body $s$-wave bound state. Coincidently, this
corresponds to the existence 
of a hyperradial potential curve, which
approaches the
corresponding (negative)
two-body binding energy in the absence of the trapping potential 
asymptotically. 
Around the pole region,
this hyperradial potential curve is
energetically close to the lowest
potential curve with gaseous-like character, and therefore, coupling 
between these two potential curves with molecular-like
and gaseous-like character, respectively, can occur.

For positive $a_{sc}$, the ``standard'' HF/GP treatment has been 
improved by including 
an additional mean-field term~\cite{braa97,fabr99,blum01}.
This modified treatment leads to an improved description
of BECs with small and medium
interaction parameters, however, exhibits divergences 
similar to that observed for the standard 
mean-field treatment as $a_{sc} \rightarrow \infty$.
Therefore the modified mean-field treatment does not overcome
the limitations of the standard mean-field treatment addressed above. 

To improve upon the GP (or equivalently, the HF) treatment,
several 
groups~\cite{drum98,timm99,timm99a,mack00,gora01,holl01,holl01a,hope01} 
developed
mean-field formalisms within the shape-independent 
contact potential approximation.
These approaches go beyond typically applied approximations
by introducing a molecular field,
and are aimed at describing condensate physics in the presence
of a Feshbach resonance.
Application of these formalisms to calculating the
energetics as a function of $a_{sc}$, i.e., in the regions where 
$a_{sc} \approx a_{cr}$ and $|a_{sc}| \rightarrow \infty$
would be useful.
The present study may then be used to benchmark such treatments, and
would thus provide a strong link between many-body and 
mean-field treatments. 

To make further contact with mean-field treatments,
comparisons between our three-body excitation frequencies
and those calculated within the 
random phase approximation can be made.
This comparison will be discussed in a future publication.

\section{Conclusions}
\label{summary} 
This paper discusses 
microscopic studies for three $^{87}$Rb atoms
interacting via a sum of simple two-body model potentials
under external confinement.
Our study is important in light of recent experiments
of atomic gases trapped in an optical lattice with an occupancy of 1-3
atoms per lattice site~\cite{grei02,stoo02}.
These experiments open the opportunity to systematically study
strongly-interacting few-body
systems, and therefore enter regimes
discussed in the present paper from a theoretical microscopic point of view.
The overall behavior of the zero temperature energy levels discussed in 
Secs.~\ref{energies} and \ref{meanfield} is independent of the exact shape of
the two-body potential, though the detailed behavior
of the energy levels and
of the hyperradial potential curves at small
hyperradii $R$ depends on the shape of
the two-body potential. The goal of this paper is to point out the
gross behavior of the energy levels 
as a function of the two-body $s$-wave scattering length 
$a_{sc}$. Therefore, we consider only one class of two-body potentials.
Our main results are derived for a relatively large trapping frequency, 
$\nu_{ho}=78$kHz.
As pointed out in Sec.~\ref{energies}, dependencies on
$\nu_{ho}$ exist, however, the gross features 
of the energy level scheme are independent of the magnitude of $\nu_{ho}$.

Our time-independent calculations, performed mostly within the adiabatic
approximation, include molecular-like and gaseous-like states,
whereas standard mean-field treatments only treat the latter states.
Our study
reveals a microscopic understanding of the decay of 
a BEC in agreement with a study by 
Bohn~{\em{et al.}}~\cite{bohn98}.
Furthermore, 
the present microscopic three-body study reveals a {\em{smooth 
non-diverging}} behavior of the energy levels around the pole
region where $|a_{sc}| \rightarrow \infty$, 
and provides a detailed picture of the 
corresponding physics through analysis of hyperspherical
potential curves.

Section~\ref{meanfield} compares our microscopic 
energy levels with those calculated at the HF level.
While mean-field theories are commonly derived in the large particle
limit, the HF equation is valid for any number of particles and density
(though it might result in a more accurate description for systems
with large $N$
and low density).
The energy level pattern for systems with more than three particles 
may differ in detail from our
three-body energy levels presented here,
however,
the overall behavior will be similar and we expect our main conclusions to 
generalize to systems with more than $N=3$ particles.

Our study raises several discussion points.
Most of all, we 
have only briefly discussed the effect of channel coupling on the 
behavior of the energy level scheme (see Fig.~\ref{energy_cc}).
To obtain detailed information
about the dynamics of the condensate, inclusion of channel coupling,
or at least a Landau-Zener-type
analysis, is required.
Such an analysis would allow the following question to be
addressed (see Fig.~\ref{fig_energy1}):
Is it possible to
form a BEC with positive $a_{sc}$ 
that resides in a hyperradial curve other than the 
lowest curve with BEC
character, and then decrease $a_{sc}$ until $a_{sc}<a_{cr}$ 
(e.g. through the
use of a Feshbach resonance)
without destroying the condensate?
Answers to this question certainly require a more
detailed analysis,
and will be left for future study.

To extend the present work
to many (more than four) particles is computationally infeasible.
We hope, however, that our results can eventually be compared 
with modified mean-field treatments that implicitly include molecular 
states~\cite{drum98,timm99,timm99a,mack00,gora01,holl01,holl01a,hope01}.
Our study may then provide
an additional link between many-body and mean-field physics.
Lastly, our work relates to 
various many-body treatments using hyperspherical
coordinates~\cite{bohn98,hart00,kim00,sore01} that neglect
molecular physics. 
(An exception is the recent study by 
S{\o}rensen~{\em{et al.}}~\cite{sore01}, which
includes molecular states explicitly.)

\acknowledgments
We thank B.~D. Esry for providing access to his B-spline code, and
J. Macek for suggesting that we apply Eqs.~(4) and (5) 
of Reference~\cite{niel99}.
This work was supported by the National Science Foundation.



\begin{figure}[tbp]
\caption{Upper panel:
Adiabatic potential curves, $U_{\nu}$,
plus trapping potential, $V_{trap}$ 
($\nu_{ho} = 78$kHz, $a_{ho}=731$a.u.),
as a function of the hyperradius $R$
for $N_b=2$ and
$a_{sc}=228$a.u.. The dashed curves ($\nu = 0,1$) describe
molecular-like states while the solid curves describe gaseous-like states.
Lower panel: Blowup of the gaseous region. In addition, dotted 
lines show
the potential curves for vanishing interaction potential $V$.
Note the logarithmic $R$ scale.
}
\label{fig_pot1}
\end{figure}

\begin{figure}[tbp]
\caption{
Internal energy levels $E_{\nu n}^{int}$ (solid lines)
as a function of 
$[\mbox{arctan}(a_{sc}/a_{ho})]/ \pi$
for three particles in a 
spherical trap 
with trapping frequency $\nu_{ho} = 78$kHz ($a_{ho}=731$a.u.).
Dotted lines indicate the ``internal'' HF energy $E_{HF}^{int}$ (see text).
The ``first cycle'' ($N_b=1$, see text) corresponds to 
$[\mbox{arctan}(a_{sc}/a_{ho})]/ \pi \in [-0.5,-1.5]$, and the ``second
cycle'' ($N_b=2$, see text) to 
$[\mbox{arctan}(a_{sc}/a_{ho})]/ \pi \in  [-1.5,-2.5]$
($[\mbox{arctan}(a_{sc}/a_{ho})]/ \pi  =-0.5,-1,-1.5,-2,-2.5$
corresponds to $|a_{sc}|=\infty,0,\infty,0,\infty$).
Labels A-F indicate systems with different 
scattering length for which Fig.~\protect\ref{fig_pot2} shows the 
corresponding adiabatic potential curves
(A: $a_{sc}= 2138$a.u.;
B: $a_{sc}= 103$a.u.;
C: $a_{sc}= 0$a.u.;
D: $a_{sc}= -31$a.u.;
E: $a_{sc}= -130$a.u.; and
F: $a_{sc}= -1970$a.u.).
}
\label{fig_energy1}
\end{figure}

\begin{figure}[tbp]
\caption{
Adiabatic potential curves, $U_{\nu}$,
plus trapping potential, $V_{trap}$, 
for three particles in a trap
with $\nu_{ho} = 78$kHz ($a_{ho}=731$a.u.)
as a function of the hyperradius $R$.
Solid lines indicate the numerical potential curves
for A: $a_{sc}= 2138$a.u.;
B: $a_{sc}= 103$a.u. ($d$-wave shape resonance);
C: $a_{sc}= 0$a.u.;
D: $a_{sc}= -31$a.u. (first signature of decay);
E: $a_{sc}= -130$a.u. (vanishing potential barrier); and
F: $a_{sc}= -1970$a.u..
These scattering lengths are also indicated in 
Fig.~\protect\ref{fig_energy1}.
The thick solid line indicates the lowest gaseous level
(panel A-D), and the ``decayed'' condensate state (panel E and F),
respectively.
In addition, dotted 
lines show
the potential curves for vanishing two-body interaction potential.
}
\label{fig_pot2}
\end{figure}

\begin{figure}[tbp]
\caption{
Internal energy levels $E_{\nu n}^{int}$ (solid lines)
as a function of 
$[\mbox{arctan}(a_{sc}/a_{ho})]/ \pi $
for three particles in a 
spherical trap 
with trapping frequency $\nu_{ho} = 780$Hz ($a_{ho}=7307$a.u.) for $N_b=2$.
Diamonds  indicate the internal energies at large negative
$a_{sc}$. 
A dotted line indicates the ``internal'' HF energy $E_{HF}^{int}$ (see text).
}
\label{fig_energy2}
\end{figure}

\begin{figure}[tbp]
\caption{
Internal adiabatic energy levels $E_{\nu n}^{int}$ 
(dotted, solid and dashed lines)
together with energies calculated using the 
coupled-adiabatic-channel approach including four channels (pluses)
as a function of 
$[\mbox{arctan}(a_{sc}/a_{ho})]/ \pi $
for three particles in a 
spherical trap 
with trapping frequency $\nu_{ho} = 78$kHz ($a_{ho}=731$a.u.).
Dotted lines indicate ``plunging'' molecular levels, while 
solid and dashed lines indicate energy levels in the lowest and in the
second lowest gaseous adiabatic potential curve, respectively.
}
\label{energy_cc}
\end{figure}

\newpage
Fig.~1\\
\centerline{\epsfxsize=6.0in\epsfbox{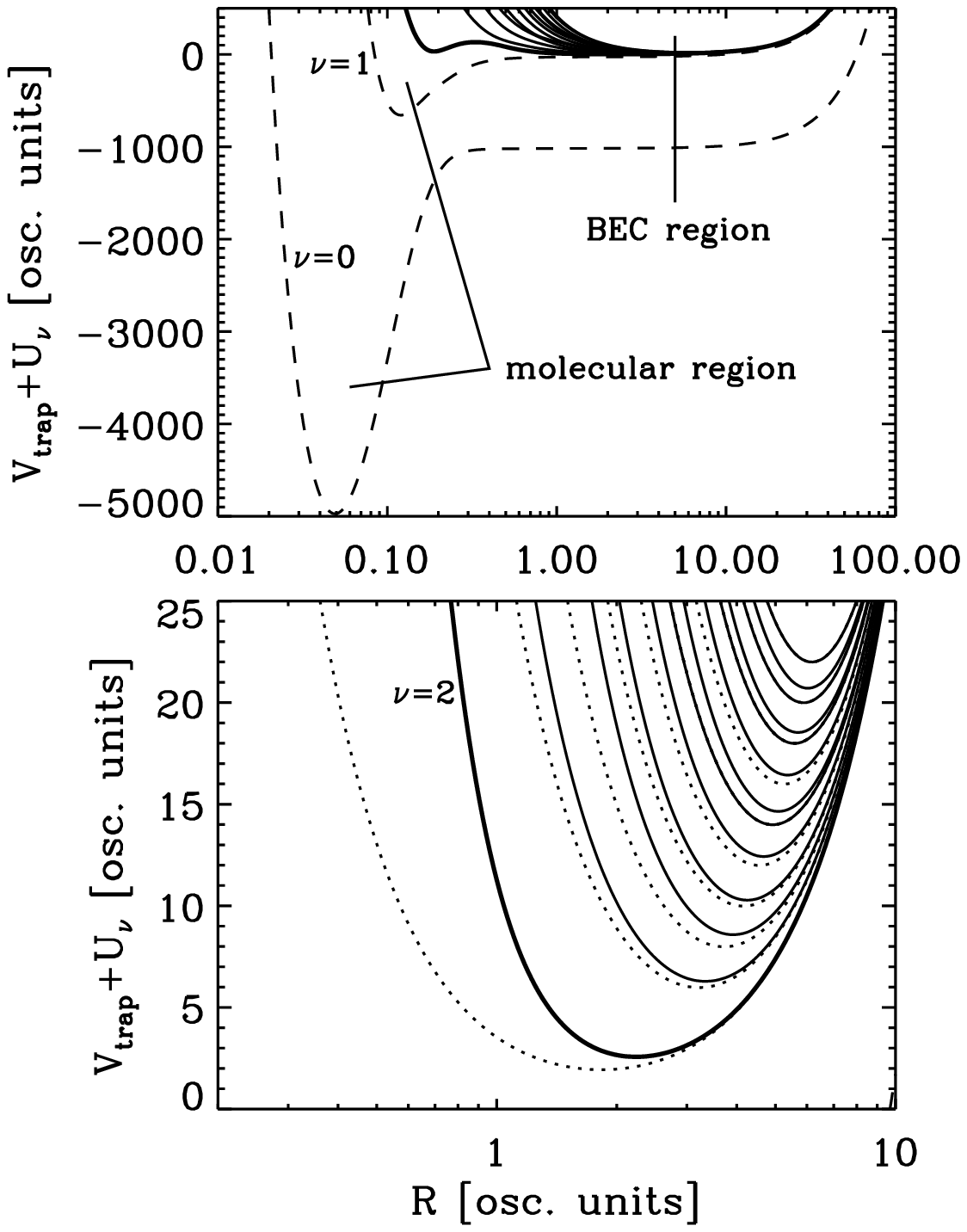}}
\newpage
Fig.~2\\
\centerline{\epsfxsize=6.0in\epsfbox{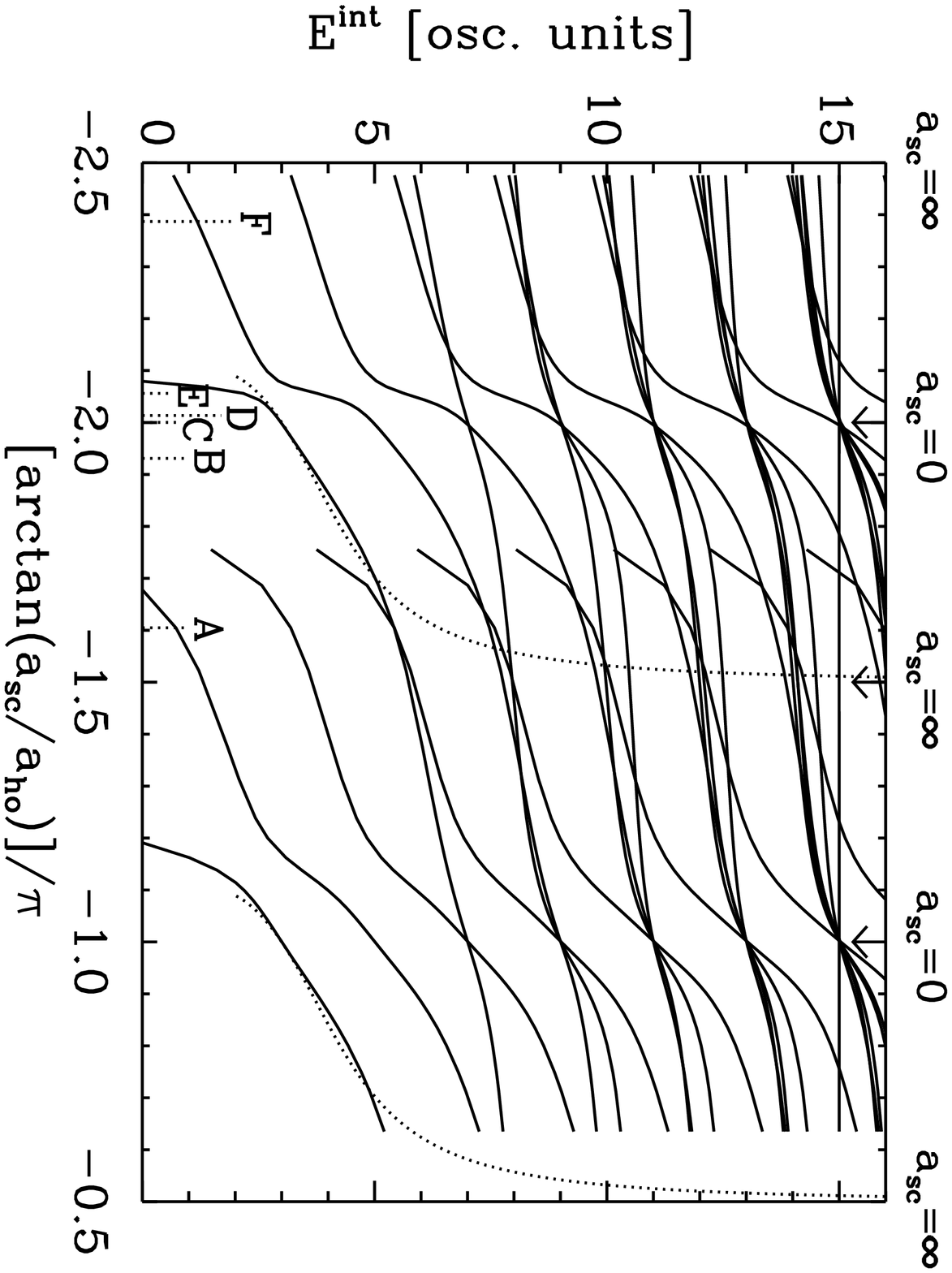}}
\newpage
Fig.~3\\
\centerline{\epsfxsize=6.0in\epsfbox{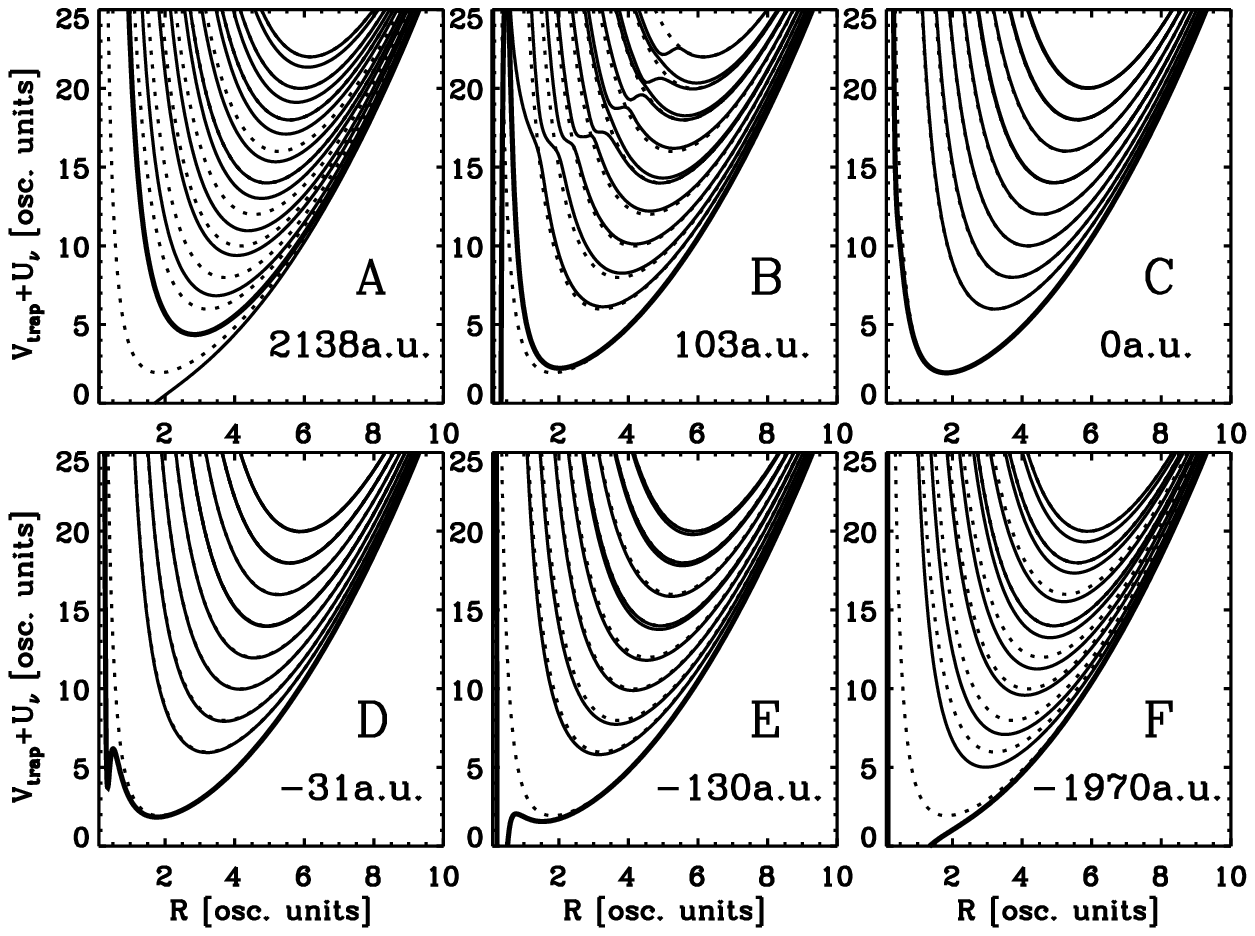}}
\newpage
Fig.~4\\
\centerline{\epsfxsize=6.0in\epsfbox{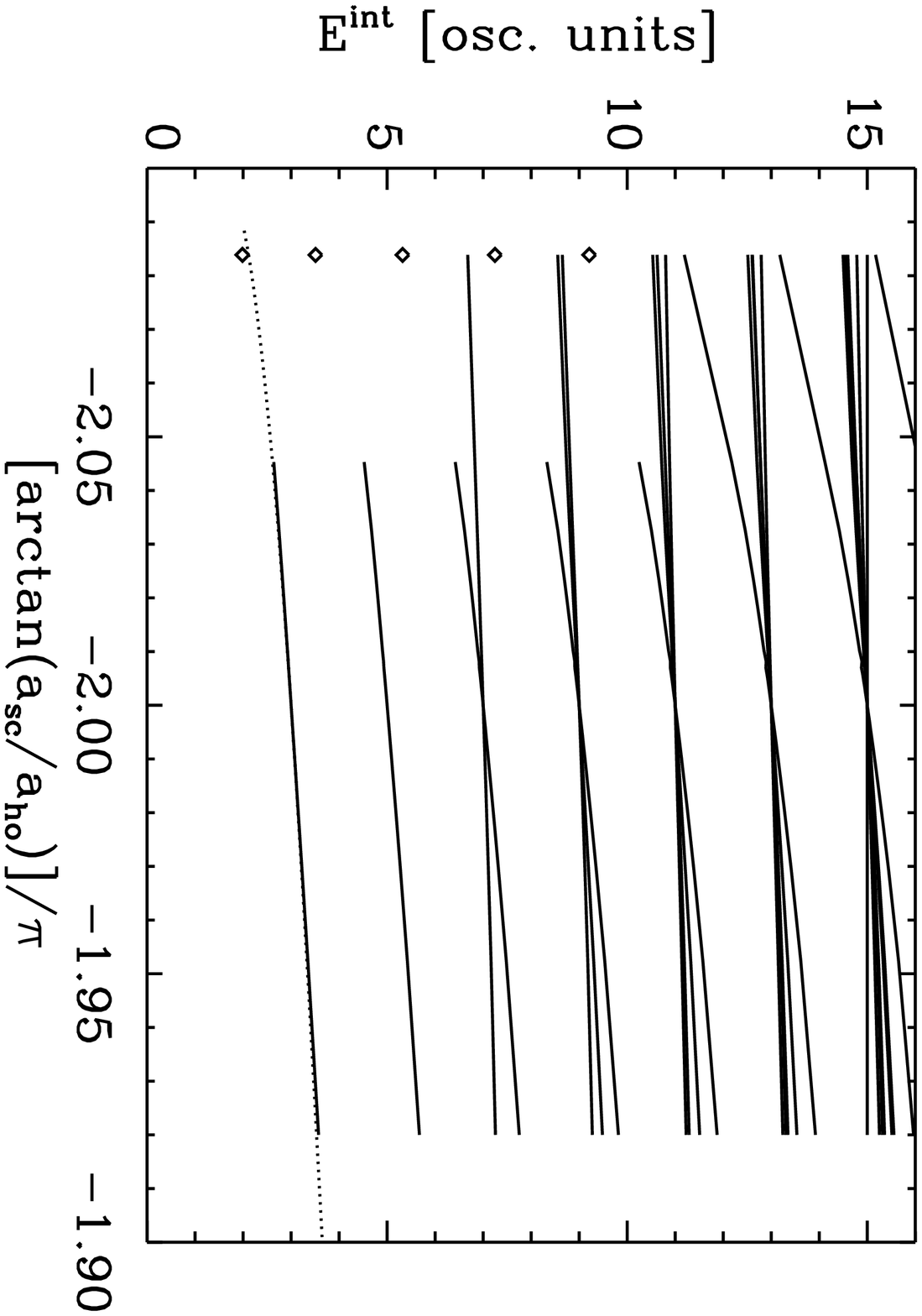}}
\newpage
Fig.~5\\
\centerline{\epsfxsize=6.0in\epsfbox{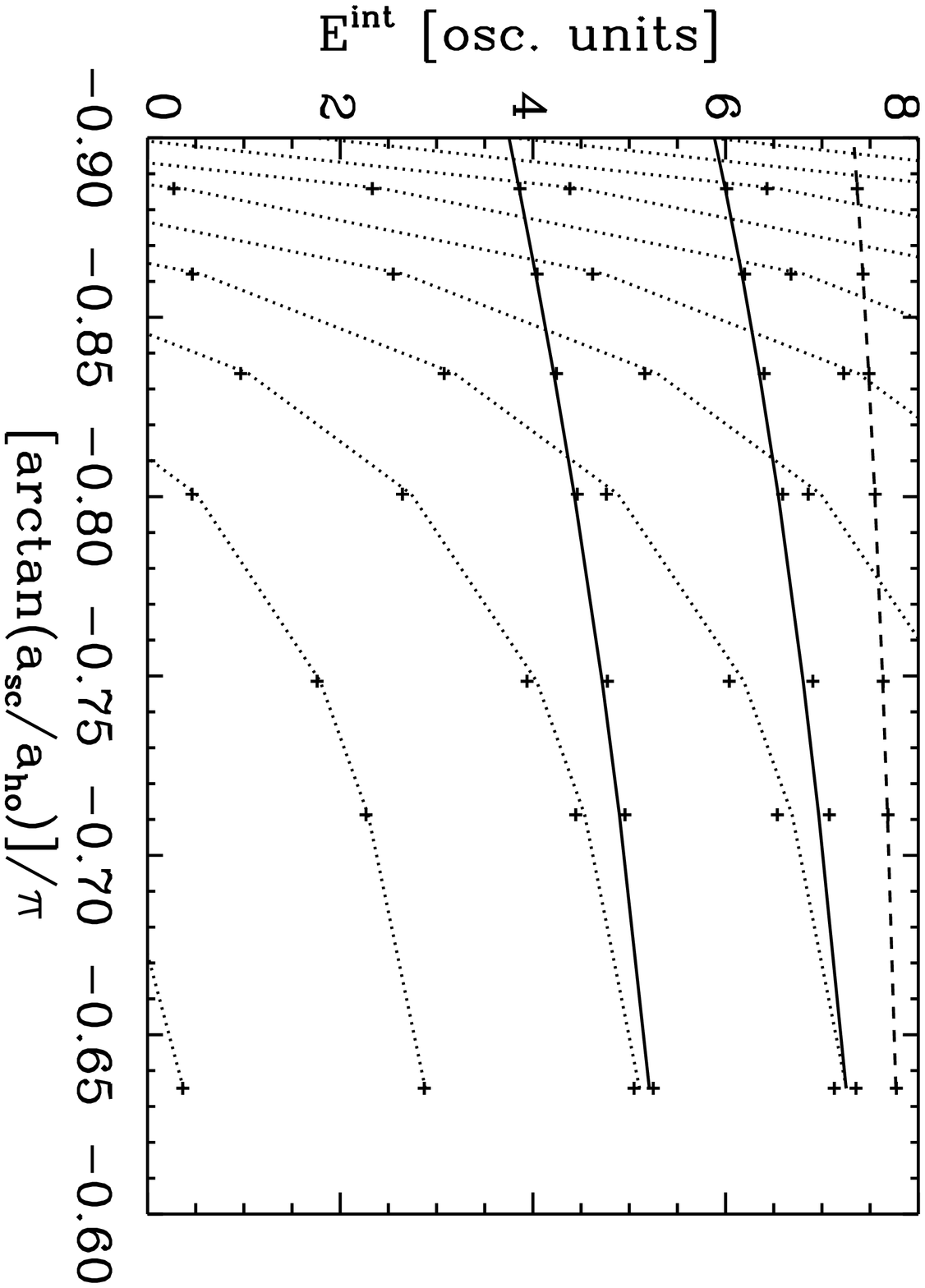}}

\end{document}